# ULTRAFAST PULSE SHAPING APPROACHES TO QUANTUM COMPUTING

DEBABRATA GOSWAMI[a]

*Indian Institute of Technology, Kanpur 208016, India*
*dgoswami@iitk.ac.in*

Quantum computing exploits the quantum-mechanical nature of matter to exist in multiple possible states simultaneously. This new approach promises to revolutionize the present form of computing. As an approach to quantum computing, we discuss ultrafast laser pulse shaping, in particular, the acousto-optic modulator based Fourier-Transform pulse-shaper, which has the ability to modulate tunable high power ultrafast laser pulses. We show that optical pulse shaping is an attractive route to quantum computing since shaped pulses can be transmitted over optical hardware and the same infrastructure can be used for computation and optical information transfer. We also address the problem of extending coherence-times for optically induced processes.

## 1 Introduction

Scaling issues in quantum computing has raised concerns on the use of current technologies for quantum computing and constant efforts are in progress towards newer approaches. One of the newer approaches involves the use of molecules interacting with lasers. Use of optical schemes is attractive as it incorporates teleportation issues also. A critical problem, however, of using molecules is in the short coherence timescales, typical of molecular states. Consequently, an immediate bottleneck in the implementation of a large-scale quantum computer lies in the fact that it is difficult to keep interacting molecular systems in the state of coherent superposition. The short coherence timescales of the molecular states are due to two important factors, one being the intermolecular interaction while the other is due to the intramolecular processes. The intermolecular coherence can be kept at a minimum with the use of molecular beam techniques; however, the intramolecular relaxation is hard to control. We discuss here a scheme to control the Intramolecular Vibrational Relaxation (IVR) through the use of optimally shaped pulses [1]. This effort of decoherence control would effectively allow access to isolated systems that behave like two-level system in spite of their multilevel structure.

Furthermore, even within the current practical situation of restricting ourselves to the use of a small interacting molecular system, where only a small number of qubits are available for computation, we show that it is possible to achieve higher compute power if systems consisting of only a few atoms or molecules acting as compute nodes are connected together with the present day fibre-optic network and are controlled and operated using ultrashort shaped optical pulses. In this approach, not only can we do ensemble quantum computing, we can also take advantage of the network bandwidth of terra-bits/sec regime.

---

[a] on leave from Tata Institute of Fundamental Research, Mumbai 400005, India

## 2  Decoherence Control

*2.1 Formalism*

The simplest model describing a molecular system is an isolated two-level system or ensemble without relaxation or inhomogeneities. Often, this is a practical model for systems interacting with femtosecond laser pulses and a linearly polarized pulse is applied to the |1>→|2> transition, where |1> and |2> represent the ground and excited eigenlevels, respectively, of the field-free Hamiltonian. We consider the case of N-photon interaction in analogy to a 1-photon interaction as shown in figure 1.

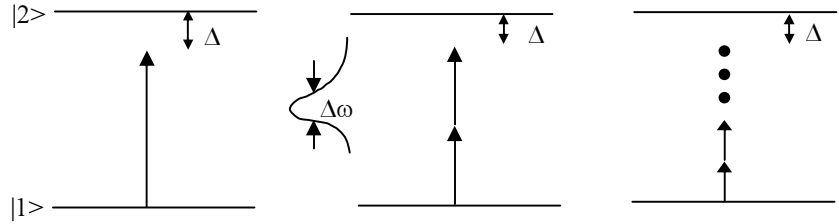

Figure 1 Schematic of single, two and multi-photon processes, respectively.  Symbols and notations defined in text.

The total laboratory-frame Hamiltonian for a two-level system under the effect of an applied laser field, in the special multiphoton case of $N^{th}$ photon allowed transition is [2]:

$$H = \begin{pmatrix} \hbar\omega_1 & \frac{1}{2k}(\mu_1.E)^N \\ \frac{1}{2k}(\mu_1.E^*)^N & \hbar\omega_2 \end{pmatrix} = \hbar \begin{pmatrix} -\frac{\omega_R}{2} & \frac{1}{2}\frac{(\mu_1.\mathcal{E})^N}{\hbar k}e^{iN(\omega t+\phi)} \\ \frac{1}{2}\frac{(\mu_1.\mathcal{E}^*)^N}{\hbar k}e^{-iN(\omega t+\phi)} & \frac{\omega_R}{2} \end{pmatrix} \quad (1)$$

where $\omega_R = \omega_2 - \omega_1$ is resonance frequency and $\hbar\omega_1, \hbar\omega_2$ are the energies of ground and excited state respectively, and $\mu_1$ is the individual transition dipole moment.  The virtual levels for two-photon (or multiphoton) case can exist anywhere within the bandwidth $\Delta\omega$ of the laser pulse (figure 1).  The proportionality constant k has dimensions of (energy)$^{N-1}$ and is unity for a 1-photon case.  This particular case of multiphoton picture allowing only $N^{th}$ photon transition is still a two-level system as the Stark effect of the multiphoton transition only adjusts the overall energy of the photon necessary to carry out the $N^{th}$ photon transition.  Since this particular form of Hamiltonian is difficult to work with, it is useful to perform a rotating-frame transformation to a frequency modulated (FM) frame, where:

$$H^{FM} = \hbar \begin{pmatrix} \Delta - N\dot{\phi}(t) & \frac{\Omega_1}{2} \\ \frac{\Omega_1^*}{2} & 0 \end{pmatrix} \quad (2)$$

This investigates off-resonance behavior of continuously modulated pulses with an additional zero-point energy shift of $(\Delta - N\dot{\phi})/2$, under assumption that the transient dipole moment of individual intermediate virtual states in the multiphoton ladder add constructively to the final state transition dipole moment ($\mu$) over N-photon electric field interaction.  The approximation is particularly valid for multiphoton interaction with femtosecond pulses where no intermediate virtual level dynamics can be observed. We define multiphoton Rabi frequency as, $\Omega_1(t)=\mu.(\varepsilon(t))^N/k\hbar$ and $\Omega^*_1(t)=\mu.(\varepsilon^*(t))^N/k\hbar$. The time derivative of the phase function $\dot{\phi}(t)$ (i.e., frequency modulation) appears as additional resonance offset besides the time-independent detuning $\Delta = \omega_R - N\omega$ (figure 1), while the direction of field in the orthogonal plane remains fixed. Time evolution of the unrelaxed two-level system can be evaluated by integrating the Liouville equation [2]:

$$\frac{d\rho(t)}{dt} = \frac{i}{\hbar}[\rho(t), H^{FM}(t)] \tag{3}$$

With this background on the two-level system, we proceed to a multilevel formalism involving IVR. In the conventional zeroth order description of intramolecular dynamics, the system can be factored into an excited state (bright state) that is radiatively coupled to the ground state, and nonradiatively to other optically inactive bath states (dark states), which follows from the optical selection rules (Figure 2 (inset)). Such dark states can belong to very different vibrational modes in the same electronic state as the bright state, or can belong to different electronic manifolds. Dark states can, however, be coupled to the bright state through anharmonic or vibronic couplings. Energy flows through these couplings and the apparent bright state population disappears. Equivalently, the oscillator strength is distributed among many eigenstates. The general multilevel Hamiltonian in the FM frame for an N-photon transition, expressed in the zero-order basis set, is:

$$H^{FM} = \hbar \begin{pmatrix} 0 & \Omega_1(t) & 0 & 0 & 0 & \cdots \\ \Omega_1^*(t) & \delta_1(t) & V_{12} & V_{13} & V_{14} & \cdots \\ 0 & V_{12} & \delta_2(t) & V_{23} & V_{24} & \cdots \\ 0 & V_{13} & V_{23} & \delta_3(t) & V_{34} & \cdots \\ 0 & V_{14} & V_{24} & V_{34} & \delta_4(t) & \cdots \\ \vdots & \vdots & \vdots & \vdots & \vdots & \end{pmatrix} \tag{4}$$

with column labels $|0\rangle, |1\rangle, |2\rangle, |3\rangle, |4\rangle, \cdots$

where, $\Omega_1(t)$ (and its complex conjugate pair, $\Omega_1^*(t)$) is the transition matrix element expressed in Rabi frequency units, between the ground state $|0\rangle$ and the excited state $|1\rangle$. The background levels $|2\rangle$, $|3\rangle$,... are coupled to $|1\rangle$ through the matrix elements $V_{12}$, $V_{23}$, etc. Both the Rabi frequency $\Omega_1(t)$ and the detuning frequency $[\delta_{1,2,\ldots} = \Delta_{1,2,\ldots} - N\dot{\phi}(t)]$ are time dependent (the time dependence is completely controlled by the experimenter).

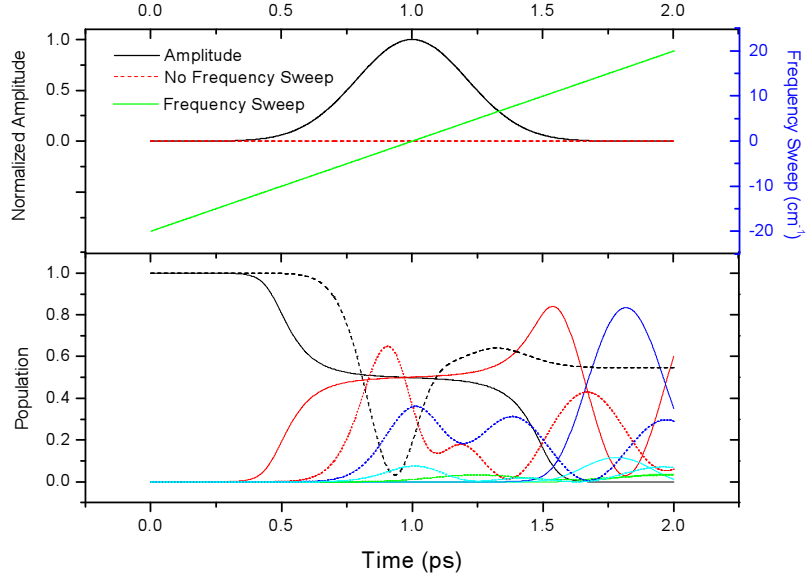

Figure 2 *(a)* Gaussian Pulse (Amplitude = $e^{-\alpha t^2}$) with no frequency sweep (dashed line, $\dot{\phi}(t) = 0$) and linear frequency sweep (solid line, $\dot{\phi}(t) \propto t$) interacts with a model Anthracene molecule. *(b)* (Inset) Schematic of IVR for anthracene molecule from Ref. [1] based on data extracted from experimental measurements in Ref. [3]. Population evolution with no sweep (dashed) and linear frequency sweep (solid). The colors represent the different states involved. Dephasing (dashed line) is suppressed over the linear frequency at the peak of the pulse (solid line).

For a multilevel system, the induced optical AC Stark-shift by the frequency swept pulse moves the off-resonant coupled levels far from the resonant state leading to an effective decoupling. Under the perfectly adiabatic condition, pulses with the even terms in the Taylor series return the system to its unperturbed condition at the end. In fact, all higher-order odd terms behave in one identical fashion and the even terms behave in another identical fashion. It is only during the pulse, that Stark-shifting of dark states is decoupled and IVR restriction is possible in the multi-level situation.

*2.2 Results*

We choose a model IVR system based on the experimental results on the fluorescence quantum beats in jet-cooled Anthracene [3]. We incorporate the respective values (in GHz) of $\Delta_{1,2,\ldots 4}$ are 3.23, 1.7, 7.57 and 3.7; and $V_{12}$=-0.28, $V_{13}$=-4.24, $V_{14}$=-1.86, $V_{23}$=0.29, $V_{24}$=1.82, $V_{34}$=0.94 in Eqn. (4), we obtain the full zero-order Hamiltonian matrix that can simulate the experimental quantum beats (figure 2b (dashed lines)) upon excitation with a transform-limited Gaussian pulse (i.e., $\dot{\phi}(t) = 0$). Since |0> and |1> do not form a closed two-level system, considerable dephasing occurs during the second half of the Gaussian pulse. Thus, in a coupled multilevel system, simple unchirped pulses cannot be used to generate sequences of π/2 and π pulses, as in NMR. The dark states start contributing to the dressed states, well before the pulse reaches its peak, and results in redistributing the population from the bright state (|1>) into the dark states (figure 2b (dashed lines)).

A linear sweep in frequency of the laser pulse can be generated by sweeping from far above resonance to far below resonance (blue to red sweeps), or it's opposite. For a sufficiently slow linear frequency sweep, the irradiated system evolves with the applied sweep and the transitions are "adiabatic". If this adiabatic process is faster than the characteristic relaxation time of the system, such a laser pulse leads to a smooth population inversion, i.e., an adiabatic rapid passage (ARP) [4]. If the frequency sweeps from below resonance to exact resonance with increasing power, and then remains constant, adiabatic *half* passage occurs and photon locking is achieved with no sudden phase shift. However, even under adiabatic *full* passage conditions, figure 2b shows that there is enough slowing down of the E field to result in photon locking over the FWHM of the pulse. These results hold even under the particular multiphoton condition where only an $N^{th}$ (N≥2) photon transition is possible [5]. Theoretically, scaling the number of dark states is possible as long as there is finite number of states and there are no physical limitations on Stark shifting.

## 3. Research Approach

Ultrafast pulse shaping [2] has taken its time to develop given the technological challenges and only now truly arbitrary pulse shapers with fast update rates have become available. A major contributor to this development has been the Fourier Transform Pulse Shapers, such as an acousto-optic modulator (AOM) based programmable pulse shaper shown in figure 3. In previous sections, we have discussed how such ultrafast pulse shaping can be adapted to achieve laser-molecule based quantum computing. In this section, we discuss how to incorporate optical networking schemes in such an approach to possibly scale up the quantum computing architecture.

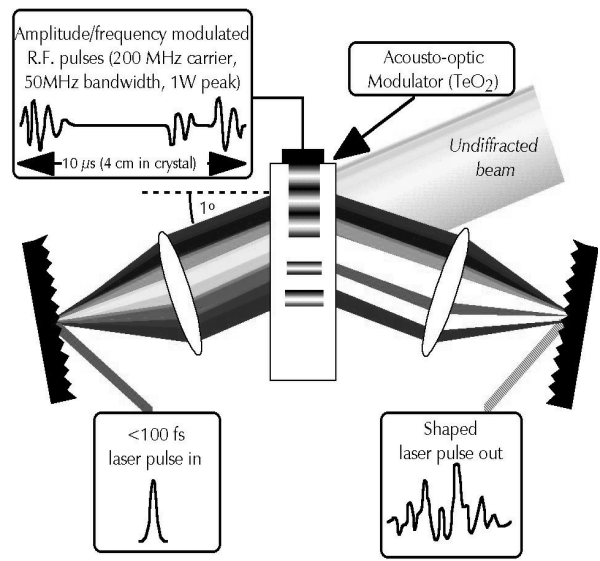

Figure 3 Schematic of an acousto-optic modulator (AOM) based ultrafast pulse shaping technology that allows arbitrary programmable pulse shaping at high repetition rate of ~$10^6$ Hz.

A typical visualization of a quantum computer network would have nodes consisting of quantum storage devices, where information can be stored for very long times either in ground or in some metastable excited states of atoms, molecules or ions. The quantum information can be transferred from one node of the network to the other using photons. The nodes would carry out the required computations and also serve as a storage or memory unit. The storage time is limited by decoherence time. Transferring quantum information between the two nodes without allowing for decoherence is very difficult. There are already some proposals in quantum communication to transmit and exchange quantum information between distant users, which includes distribution of quantum secure key information for secure communication. Teleportation allows an arbitrary unknown quantum state to be conveyed from one distant part to another with perfect fidelity by the establishment of a maximal entangled state of two distant quantum bits. However, the bottleneck for communication between distant users is the scaling of the error probability with the length of the channel connecting the users. The error results from amplitude and phase damping.

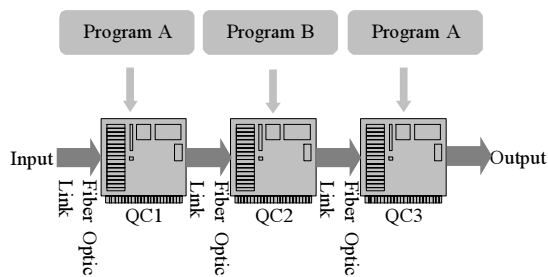

Figure 4 Schematic of the AOM pulse shaper based fiber- optic networked distributed quantum-computing framework as discussed in the text.  For clarity purposes we show the schematic with only three quantum computers (QC).

Our proposal in realizing such schemes would achieve quantum computing at nodes of the quantum networks and send photons through free space or through standard optical fibres for networking (figure 4). In such a scheme we would exploit the low cross-talk between two light signals, capacity to carry enormous amount of information encoded on the optical pulse, combined with its ability to induce the desired transitions in the target systems. Ultrafast optically shaped pulses enable us to get to our goals. Appropriately shaped optical pulse has been successfully used to induce selective excitation to control the yield of the desired product at the end of the chemical reaction either under single or multi-photon condition [2]. As discussed in the previous section, such designed pulses also have potential to control decoherence in large molecules by controlling intra-molecular vibrational relaxation. The quantum interaction with the incident shaped pulse is in effect as long as the pulse is present irrespective of the dephasing mechanism. This would add to the high fidelity of the systems, since the presence of the environment, which would otherwise destroy the coherence, would not affect the operation. Shaped optical pulses have furthermore been used to demonstrate the quantum logic gate (CNOT) in such systems [1].

The shaped pulses can be split into a number of different parts which can carry different train of pulses at different timescales. This would provide leverage to control the various nodes where molecular systems interact with shaped pulses to carry out various instructions and perform quantum computing activity at each node during the pulse duration. This could enable the processing of different quantum computational steps at various nodes simultaneously, which implies that we can *paralleliz*e our code and distribute the task over the network. At the end of the computation we can read out the results by sending in a "read pulse" and recombining the results. Essentially, this is distributed quantum computing over the network using shaped pulses. Currently, $10^6$ bits can be transmitted/encoded in a single burst of light [6] with the present day optical pulse shaping technology as shown in Fig. 3. The repetition rate from the laser source is about 50 to 100 MHz. Thus, one would be able to use terabit/sec bit of communication channel through the existing infrastructure available with the optical community. Once such a quantum computer is available at remote site, these packets acting as "quantum software" can be transferred through high-speed communication channels. Thus, it is possible to carry out quantum computation at a remote distance with the proposed scheme of shaped pulses for terabit/sec communication and molecular control (figure 4).

An additional feature of the proposed architecture could be in making use of the quantum computer as a *mainframe device*, which is connected to various other *request handlers* for quantum computational task through different optical fiber channel. The quantum computer sits at the main terminal node with requests coming from various sources, which could be handled on the time-sharing basis. Our pulse-shaping approach can pack various requests from various sources using Time-Division Multiplexing (TDM), which would add to the advantage of having shaped pulses for controlling dephasing and executing quantum gate operation. We can have one master pulse going in to *ready* the quantum computer to execute incoming quantum operation. The next set of instruction pulses would carry out its tasks and an optical pulse can read out the result. This would be then decoded and send back to the respective users. The request comes in the form of shaped pulses to send in the appropriate instructions.

Our proposed approach takes advantage of the high data transfer rate of terabit/sec communication, which is possible with the ultrafast pulse shaping technology and we should be able to perform distributed quantum computation as well as "Unix kind of multitasking environment". Earlier in an article the need to download quantum software at the end of each computation was shown [7], because of the fragile nature of the systems as well as the fundamental limitations posed by the very nature of the system. Here we do not require to teleport or transport a quantum system between the terminal ends of the network through communication channels where there is always a high chance that it might get corrupted or lost. Also this mode of transfer of transfer of the quantum systems calls for a specific kind of infrastructure to communicate between the different nodes of the network.

## 4. Epilogue

What physical systems are potentially good candidates for handling quantum information? The key concept in understanding this is the notion of quantum noise or decoherence that corrupts desired evolution of the system. The length of the longest possible quantum computation is roughly given by the ratio of $\tau_q$ (the time for which a system remains quantum-mechanically coherent) to $\tau_{op}$ (the time it takes to perform elementary unitary transformations, which involve at least two qubits). These two times are actually related to each other in many systems, since they are both determined by the strength of coupling of the system to the external world. Nevertheless, $\lambda = \tau_{op}/\tau_q$ can vary over surprisingly wide range in the kind of systems where quantum computation has been achieved [8].

## Acknowledgements

Partial support for this work has been provided by generous funding from the Ministry of Communication and Information Technology, Govt. of India.